\documentclass[prd,superscriptaddress,nofootinbib,amsmath,amssymb,aps,11pt]{revtex4}

\usepackage{bm}
\usepackage{amsfonts}
\usepackage{latexsym}
\usepackage[latin1]{inputenc}
\usepackage{graphicx}
\usepackage{amsmath}
\usepackage{palatino}
\usepackage{mathpazo}
\usepackage{booktabs}
\usepackage{dcolumn}

\def\jnl@style{\it}
\def\aaref@jnl#1{{\jnl@style#1}}

\def\aaref@jnl#1{{\jnl@style#1}}

\def\aj{\aaref@jnl{AJ}}                   
\def\apj{\aaref@jnl{ApJ}}                 
\def\apjl{\aaref@jnl{ApJ}}                
\def\apjs{\aaref@jnl{ApJS}}               
\def\apss{\aaref@jnl{Ap\&SS}}             
\def\aap{\aaref@jnl{A\&A}}                
\def\aapr{\aaref@jnl{A\&A~Rev.}}          
\def\aaps{\aaref@jnl{A\&AS}}              
\def\mnras{\aaref@jnl{Mon.~Not.~Roy.~Astron.~Soc.}}             
\def\prd{\aaref@jnl{Phys.~Rev.~D}}        
\def\prc{\aaref@jnl{Phys.~Rev.~C}}  
\def\prl{\aaref@jnl{Phys.~Rev.~Lett.}}    
\def\qjras{\aaref@jnl{QJRAS}}             
\def\skytel{\aaref@jnl{S\&T}}             
\def\ssr{\aaref@jnl{Space~Sci.~Rev.}}     
\def\zap{\aaref@jnl{ZAp}}                 
\def\nat{\aaref@jnl{Nature}}              
\def\aplett{\aaref@jnl{Astrophys.~Lett.}} 
\def\apspr{\aaref@jnl{Astrophys.~Space~Phys.~Res.}} 
\def\physrep{\aaref@jnl{Phys.~Rep.}}      
\def\physscr{\aaref@jnl{Phys.~Scr}}       
\def\commat{\aaref@jnl{Comm.~Math.~Phys.}}              
\def\science{\aaref@jnl{Science}}               
\def\cqg{\aaref@jnl{Classical Quant.~Grav.}}            
\def\jpcs{\aaref@jnl{JPCS}}                                     
\def\ijmpd{\aaref@jnl{Int.~J.~Mod.~Phys.~D}}                    
\def\grg{\aaref@jnl{Gen.~Relat.~Gravit.}}               
\def\rpp{\aaref@jnl{Rep.~Prog.~Phys.}}          
\def\npa{\aaref@jnl{Nucl.~Phys.~A}}        
\def\lrr{\aaref@jnl{Living Rev.~Rel.}}                   
\def\jcap{\aaref@jnl{J.~Cosmology Astropart.~Phys.}}    
\def\rmp{\aaref@jnl{Rev.~Mod.~Phys.}}   

\allowdisplaybreaks[1]

\begin{document}

\title{Relativistic stars in 4D Einstein-Gauss-Bonnet gravity }

\author{Daniela D. Doneva}
\email{daniela.doneva@uni-tuebingen.de}
\affiliation{Theoretical Astrophysics, Eberhard Karls University of T\"ubingen, T\"ubingen 72076, Germany}
\affiliation{INRNE - Bulgarian Academy of Sciences, 1784  Sofia, Bulgaria}

\author{Stoytcho S. Yazadjiev}
\email{yazad@phys.uni-sofia.bg}
\affiliation{Theoretical Astrophysics, Eberhard Karls University of T\"ubingen, T\"ubingen 72076, Germany}
\affiliation{Department of Theoretical Physics, Faculty of Physics, Sofia University, Sofia 1164, Bulgaria}
\affiliation{Institute of Mathematics and Informatics, 	Bulgarian Academy of Sciences, 	Acad. G. Bonchev St. 8, Sofia 1113, Bulgaria}


\begin{abstract}
In the present paper we investigate the structure of relativistic stars in 4D Einstein-Gauss-Bonnet gravity.  The mass-radius relations are obtained for realistic hadronic and for strange quark star equations of state, and for a wide range of the Gauss-Bonnet coupling parameter $\alpha$. Even though the deviations from general relativity for nonzero values of $\alpha$ can be large enough, they are still comparable with the variations due to different modern realistic equations of state if we restrict ourselves to moderate value of $\alpha$. That is why the current observations of the neutron star masses and radii alone can not impose stringent constraints on the value of the parameter $\alpha$. Nevertheless some rough constraints on $\alpha$ can be put. The existence of stable stellar mass black holes imposes $\sqrt{\alpha}\lesssim 2.6 {\rm km}$  for $\alpha>0$ while the requirement that the maximum neutron star mass should be greater than two solar masses  gives $\sqrt{|\alpha|}\lesssim 3.9 {\rm km}$  for $\alpha<0$.  We also present an exact solution describing the structure of relativistic stars with uniform energy density  in 4D Einstein-Gauss-Bonnet gravity.
\end{abstract}

\maketitle

\section{Introduction}

A new covariant modified theory of gravity in $D = 4$ space-time
dimensions which propagates only the massless graviton and bypasses the well-known Lovelock's theorem \cite{Lovelock_1971}, was recently proposed in \cite{Glavan_2020}. This theory is called 4D Einstein-Gauss-Bonnet (EGB) gravity . The theory  is first formulated in $D >4$ dimensions and its action consists of the standard Einstein-Hilbert term (with a cosmological constant) and  the Gauss-Bonnet term. The $4D$ theory is obtained as a ``dimensional regularization'' in the limit $D\to 4$ of the higher dimensional theory. It should be mentioned that 
prior to  \cite{Glavan_2020} dimensional regularization of this kind was proposed in \cite{Tomozawa_2011}. In the described  singular limit the Gauss-Bonnet invariant gives  non-trivial contributions to gravitational dynamics, while preserving the number of graviton degrees of freedom and being free from Ostrogradsky instability. 

The EGB gravity posses some interesting and attractive features. Particularly, it predicts new static and spherically symmetric  black hole solutions which differ from the well-known Schwarzschild black hole in general relativity (GR). A very interesting fact is that, although  the EGB gravity is a pure classical model, its black holes solutions formally coincide with the black holes solutions previously found in gravity with quantum corrections \cite{Cognola_2013}. Amongst the other attractive properties of the model proposed in \cite{Glavan_2020} is that it might resolve some singularity issues. For example, within the novel static and spherically symmetric black hole solutions  the gravitational force is repulsive at small distances and
thus an infalling particle never reaches the singularity. Some aspects of the novel model of 
\cite{Glavan_2020} have already attracted attention. Stability, quasinormal modes and shadow of novel black holes were studied 
in \cite{Konoplya_2020}, while  the innermost circular orbits were considered in \cite{Guo_2020}. Charged black holes with anti-de Sitter and de Sitter  asymptotics  were found in \cite{Fernandes_2020}. Asymptotically flat black hole solutions in the EGB gravity were found in \cite{Wei_2020} and \cite{Kumar_2020} and their shadows were also analyzed.

I the present paper we address the important problem of the existence of relativistic stars in the 4D Einstein-Gauss-Bonnet gravity and
their basic properties. The paper is organized as follows. In Sec. II we present the dimensionally reduced field equations describing the structure
of the relativistic stars in  EGB gravity. The analytical solution for constant density relativistic stars in EGB gravity is presented in Sec. II. The numerical solutions describing neutron stars with realistic hadronic and strange quark star equations of state are discussed in Sec. IV. The paper ends with Conclusions. 
 
\section{Basic equations and setting the problem}

We start with a $D$-dimensional Gauss-Bonnet theory minimally coupled to matter fields whose action is given by 
\begin{eqnarray}
S_D=\frac{1}{16\pi G}\int d^Dx \sqrt{-g}\left[R^{(D)} + \alpha_{*} {\cal G}^{(D)} \right] + S_{\text{matter}} .
\end{eqnarray}  
Here $R^{(D)}$ is the Ricci scalar curvature and ${\cal G}^{(D)}$ is the Gauss-Bonnet invariant  associated with the $D$-dimensional spacetime.
The parameter $\alpha_{*}$ is the Gauss-Bonnet coupling parameter.  $S_{\text{matter}}$ is the action of the matter fields. In the present paper the matter content we shall consider is a perfect fluid described by the energy density $\rho$, pressure $p$ and the $D$-velocity $u^{\mu}$. 
The energy density and pressure are related with the equation of state $p=p(\rho)$.

In the present paper we shall consider only static and spherically symmetric spacetimes and fluid configurations. The $D$-dimensional metric
is then 

\begin{eqnarray}
ds^2_{D}= - e^{2\Phi(r)}dt^2 + e^{2\Lambda(r)}dr^2 + r^{2}d\Omega_{D-2}^2,  
\end{eqnarray} 
where $d\Omega_{D-2}^2$ is the metric on the unit $(D-2)$-dimensional sphere. Under our assumptions the $D$-dimensional perfect fluid is 
characterized with $\rho=\rho(r)$, $p=p(r)=p(\rho(r))$ and $u_{t}=e^{\Phi(r)}$ as all the spacial  components $u^{i}$ of the $D$-velocity vanish.

As in \cite{Glavan_2020} the effective 4D theory is obtained in the singular limit $D\to 4$ with $\alpha_{*}=\alpha/(D-4)$.  Carefully performing the limit $D\to 4$ we find the following dimensionally reduced field equations
\begin{eqnarray}\label{DRE1}
&&\frac{2}{r} \frac{d\Lambda}{dr}=
 \, e^{2\Lambda} \, \, \frac{8\pi G\rho - \frac{(1-e^{-2\Lambda})}{r^2}[1- \alpha \frac{(1-e^{-2\Lambda})}{r^2}]}{[1 + 2\alpha \frac{(1-e^{-2\Lambda})}{r^2}]}, \\ \notag \\
&& \frac{2}{r} \frac{d\Phi}{dr}=
  e^{2\Lambda} \, \, \frac{8\pi G p + \frac{(1-e^{-2\Lambda})}{r^2}[1- \alpha \frac{(1-e^{-2\Lambda})}{r^2}]}{[1 + 2\alpha \frac{(1-e^{-2\Lambda})}{r^2}]}, \\
&& \frac{dp}{dr}= - (\rho + p) \frac{d\Phi}{dr}.  \label{DRE3}
\end{eqnarray} 
Asymptotic flatness imposes $\Phi(\infty)=\Lambda(\infty)=0$ while  the regularity at the center requires $\Lambda(0)=~0$ . 

The above system of equations (\ref{DRE1})--(\ref{DRE3}) supplemented with the equation of state of the baryonic matter $p= p(\rho)$, with the above mentioned asymptotic and regularity conditions as  well as with a specified central energy density ${\rho}_c$, describes the structure of the neutron stars in the model under consideration.

\section{Exact solution for relativistic stars with uniform energy density}

In this section we will discuss an analytical solution to the field equations (\ref{DRE1})--(\ref{DRE3})  for a relativistic  star with  uniform energy density. In oder words we shall present a generalization of the famous interior Schwarzschild solution. It is well-known that the  interior Schwarzschild solution is not completely realistic, however it describes qualitatively very well the general case of a static,
spherically symmetric perfect fluid star in general relativity and in particular, it predicts the existence of an upper limit for the stellar compactness, known in the general case as the Buchdahl inequality \cite{Buchdahl_59}. 
      
As we mentioned we consider  a star with  uniform energy density $\rho=const$. In this case the equation for $\Lambda$ is separated from
the other equations. In order to solve this equation it is convenient to introduce the new function $\zeta$ defined by

\begin{eqnarray}
\zeta= \frac{1-e^{-2\Lambda}}{r^2}.
\end{eqnarray} 
In terms of $\zeta$ the equation for $\Lambda$ takes the form

\begin{eqnarray}
\frac{d}{dr}\left[r^3(\zeta + \alpha\zeta^2)\right]=8\pi\rho r^2 .
\end{eqnarray} 
This equation can  easily  be integrated and the solution for both $\alpha>0$ and $\alpha<0$ which  is regular\footnote{We want the metric and the curvature invariants to be regular at the center of the star.} at the center of the star is  $(\zeta + \alpha\zeta^2)=\frac{8\pi \rho}{3}$. The solution of this quadratic algebraic equation which agrees with the asymptotically flat exterior solution is   

\begin{eqnarray}
\zeta= \frac{\sqrt{1 + \frac{32\pi\alpha\rho}{3}}-1}{2\alpha} .
\end{eqnarray} 
In other words we have

\begin{eqnarray}
e^{-2\Lambda}= 1- \frac{r^2}{2\alpha} \left(\sqrt{1 + \frac{32\pi\alpha\rho}{3}}-1\right)  .
\end{eqnarray} 
Having once the explicit form of $\zeta$ (equivalently of  $e^{-2\Lambda}$) we can get the metric function $\Phi$ and the pressure $p$. After long but straightforward calculations we  find 

\begin{eqnarray}
&&e^{2\Phi}= \frac{1}{4}\left(\frac{1-\alpha\zeta}{1+2\alpha\zeta}\right)^2 \left[3 \frac{1+\alpha\zeta}{1-\alpha\zeta}\sqrt{1-\zeta R^2} 
- \sqrt{1-\zeta r^2}\right]^2 , \\ \notag \\
&&p= \rho \frac{\sqrt{1-\zeta r^2} -\sqrt{1-\zeta R^2}}{3 \frac{1+\alpha\zeta}{1-\alpha\zeta}\sqrt{1-\zeta R^2} 
	- \sqrt{1-\zeta r^2}} .
\end{eqnarray} 
Here $R$ is an integration constant. It is not difficult to see that the solution has a well-defined boundary where the pressure vanishes, $p(R)=0$. Therefore, $R$ is the radius of the star. In the limit $\alpha \to 0$, as one can easily see, our solution reduces to  the Schwarzschild interior solution.  In order for the solution to be regular everywhere for $0\le r\le R$ the following inequality should be satisfied 
\begin{eqnarray}\label{Inq1}
3 \frac{1+\alpha\zeta}{1-\alpha\zeta}\sqrt{1-\zeta R^2} > 1.
\end{eqnarray}
The interior metric has to match continuously the exterior metric  given by the metric functions
\begin{eqnarray}\label{extsol}
e^{-2\Lambda_e}=e^{2\Phi_e}= 1- \frac{r^2}{2\alpha}\left[\sqrt{1+ \frac{8M\alpha}{r^3} }- 1 \right],   
\end{eqnarray}
where $M$ is the mass of the star. The matching conditions give the relation among the mass, the radius and the energy density of the star, namely 
\begin{eqnarray}
M=\frac{4\pi}{3}\rho R^3, 
\end{eqnarray}  
which in turn implies that 
\begin{eqnarray}
\zeta = \frac{\sqrt{1 + \frac{8M\alpha}{R^3}}- 1}{2\alpha}. 
\end{eqnarray} 

Let us return to the inequality (\ref{Inq1}). In the pure general relativistic case, i.e. for $\alpha=0$, this inequality is in fact the   
the Buchdahl inequality $M/R<4/9$. Our inequality (\ref{Inq1}) plays similar role, it imposes a nonlinear constraint on the mass and the radius
for fixed parameter $\alpha$ which ensures the existence of a regular star.  

\section{Numerical solutions}
In this section we will present the results for the numerical solutions describing neutron stars and strange stars in EGB gravity. We employ several realistic hadronic matter EOS presented in the piecewise approximation form \cite{Read:2008iy} while for the strange star EOS we use \cite{GondekRosinska:2008nf}. Most of the results presented below are for the APR4 EOS \cite{APR4} and the strange star SQS40 EOS \cite{GondekRosinska:2008nf}, but as discuss below, we use other EOS as well in order to impose constraints on the parameter $\alpha$ coming from the neutron star observations.

The reduced field equations (\ref{DRE1})--(\ref{DRE3}) together with the boundary conditions are solved using a fourth order Runge-Kutta method. The obtained solutions are counterchecked again the analytic outer solution \eqref{extsol} and the results show very good agreement.

The neutron star mass is obtained from the asymptotic of the metric function $\Lambda$, namely we require that very far outside the star $\exp(-2\Lambda)\approx 1-\frac{2M}{r}$ as follows from the exterior solution (\ref{extsol}). The radius of the star is obtained from the requirement that $p(R)=0$ while the baryon mass is calculated using the following integral
\begin{equation}
M_0 = \int_{0}^{R}{4\pi \rho_0 e^{\Lambda}r^2 dr}.
\end{equation}
Here $\rho_0$ is the rest mass density.

Before presenting our numerical results let us put rough constraints   on the Gauss-Bonnet coupling parameter $\alpha$ in the case when $\alpha>0$. Static, spherically symmetric and vacuum solutions to the equation of 4D EGB gravity  with mass $M$ describe black holes only 
when $\alpha\le M^2$ \cite{Glavan_2020}. This can be also seen from the exterior solution (\ref{extsol}). Having such a condition for $\alpha$, the very existence of stellar mass black holes can impose constraints on this parameter. Most probably, the 2MASS J05215658+4359220 contains the lowest mass black hole with  $M\approx 3.3 M_{\odot}$ \cite{Thompson_2019}. Then the condition  $\alpha\le M^2$ gives an upper bound $\sqrt{\alpha} \lesssim 4.9 {\rm km}$. However, according to \cite{Konoplya_2020} 
the black holes are stable for $\alpha/M^2 \lesssim 0.3$ (in our notations) which reduces the upper bound to $\sqrt{\alpha} \lesssim 2.6 {\rm km}$. In the same way, taking into account
that the black holes in EGB gravity with $\alpha<0$  are stable for  $|\alpha|/M^2 \lesssim 4$ (in our notations) \cite{Konoplya_2020},   we  find $\sqrt{|\alpha|}\lesssim 9.7 {\rm km}$.
However, as we show below, this estimate can be improved by using our numerical results for the neutron stars.

In the numerical results, presented in this section, we use the dimensionless parameter $\alpha \to \alpha/R^2_{0} $ where $R_{0}=1.47664 km$
corresponds to one half of the solar gravitational radius. In terms of the dimensionless parameter $\alpha$ the above upper bounds are given by $\alpha \lesssim 3.2$ and $|\alpha|\lesssim 43.6$.

\begin{figure}
	\includegraphics[width=0.45\textwidth]{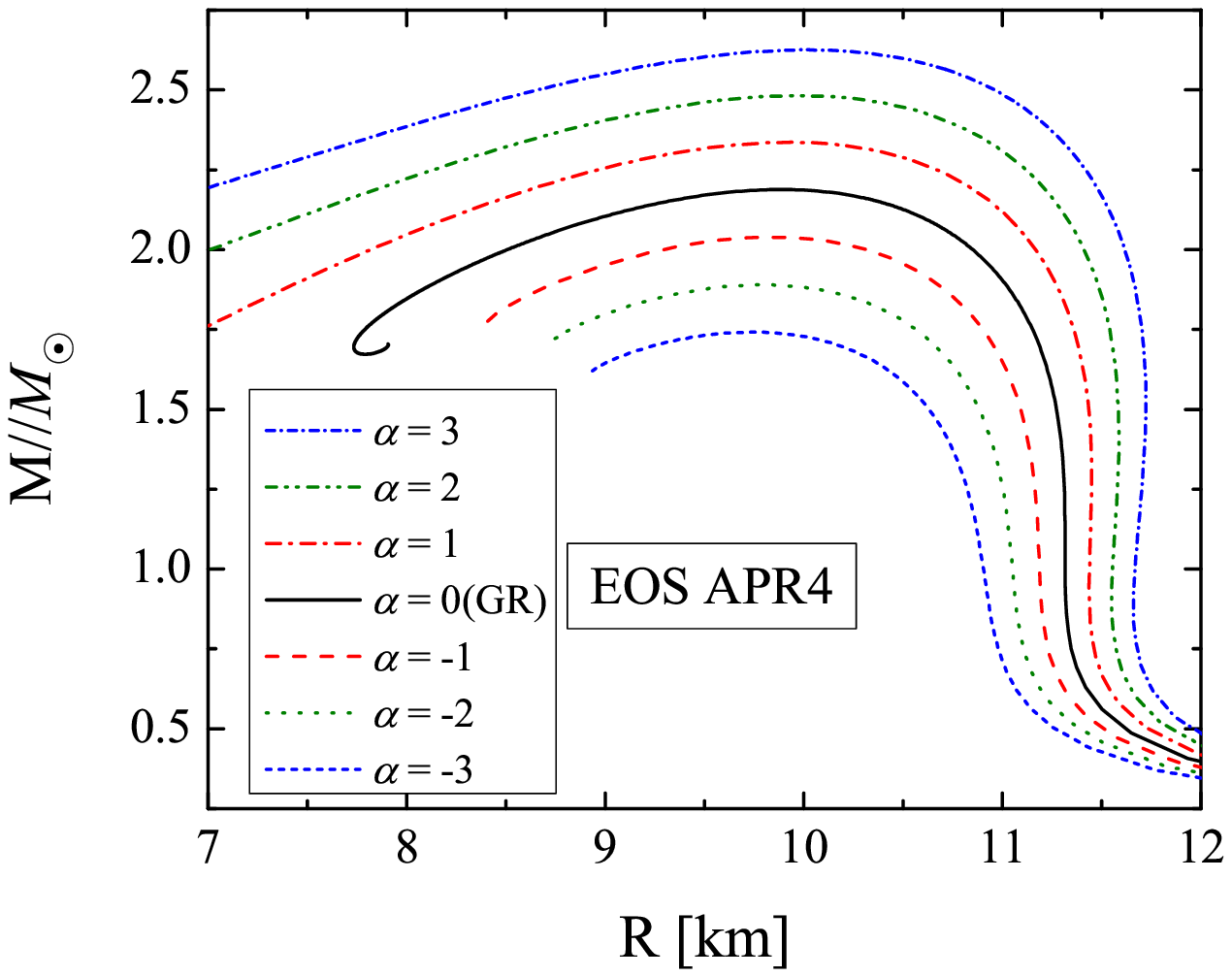}
	\includegraphics[width=0.45\textwidth]{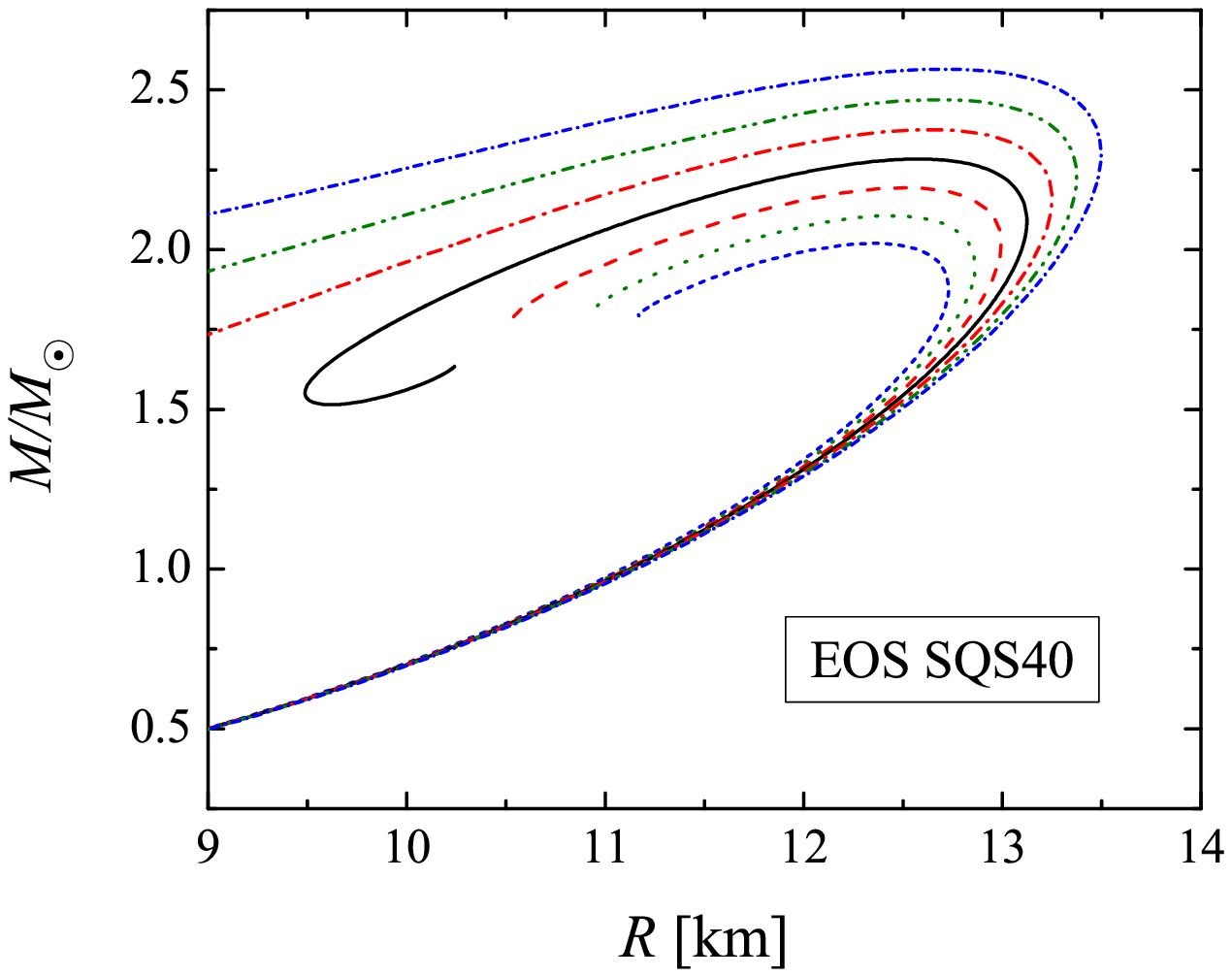}
	\caption{The mass as a function of radius for different values of the parameter $\alpha$ and two equations of state -- the hadronic matter APR4 EOR (left panel), and the strange quark star SQS40 EOS (right panel). The color and style coding of lines for different $\alpha$ is the same in both panels.}
	\label{fig:MR}
\end{figure}

\begin{figure}
	\includegraphics[width=0.45\textwidth]{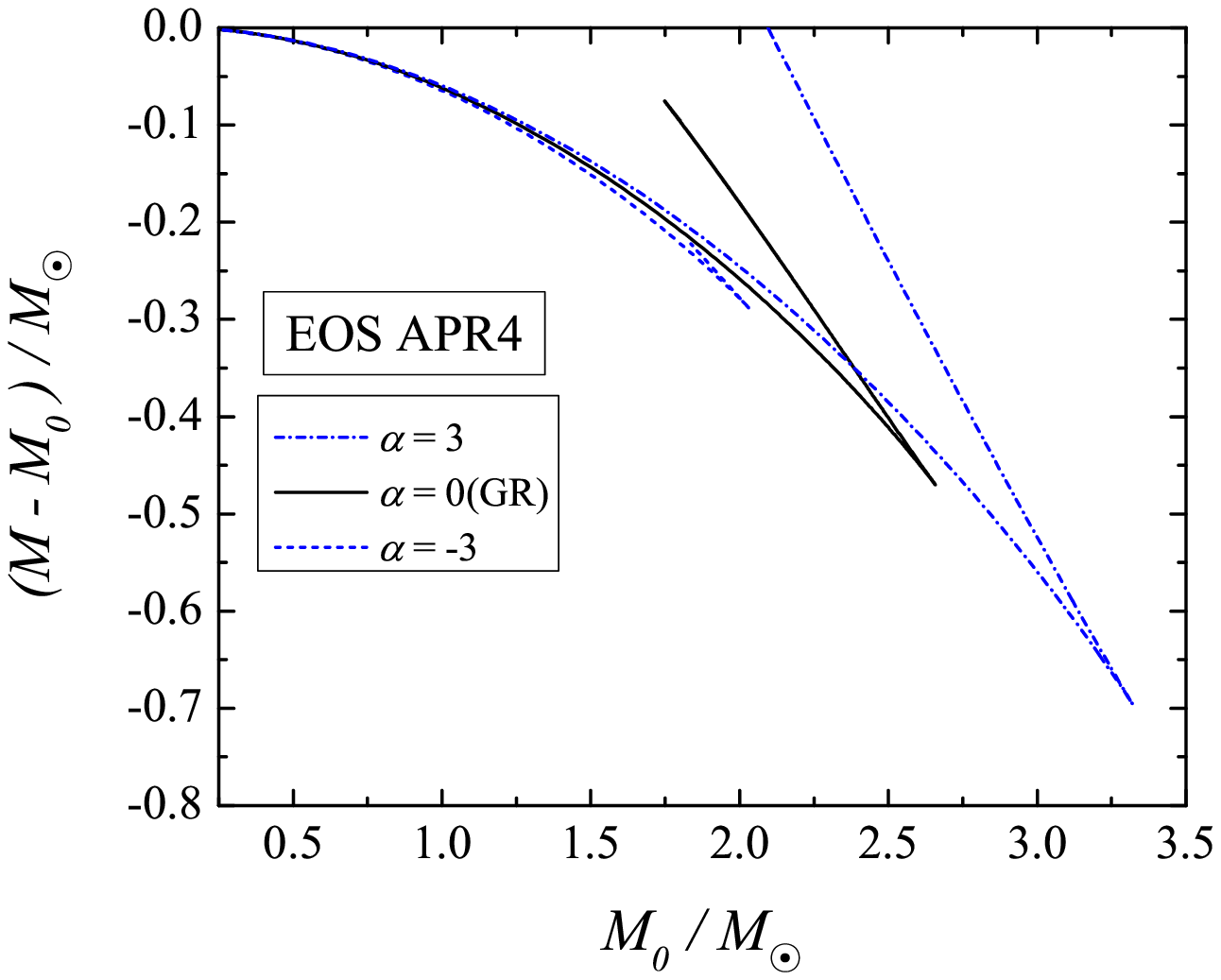}
	\includegraphics[width=0.45\textwidth]{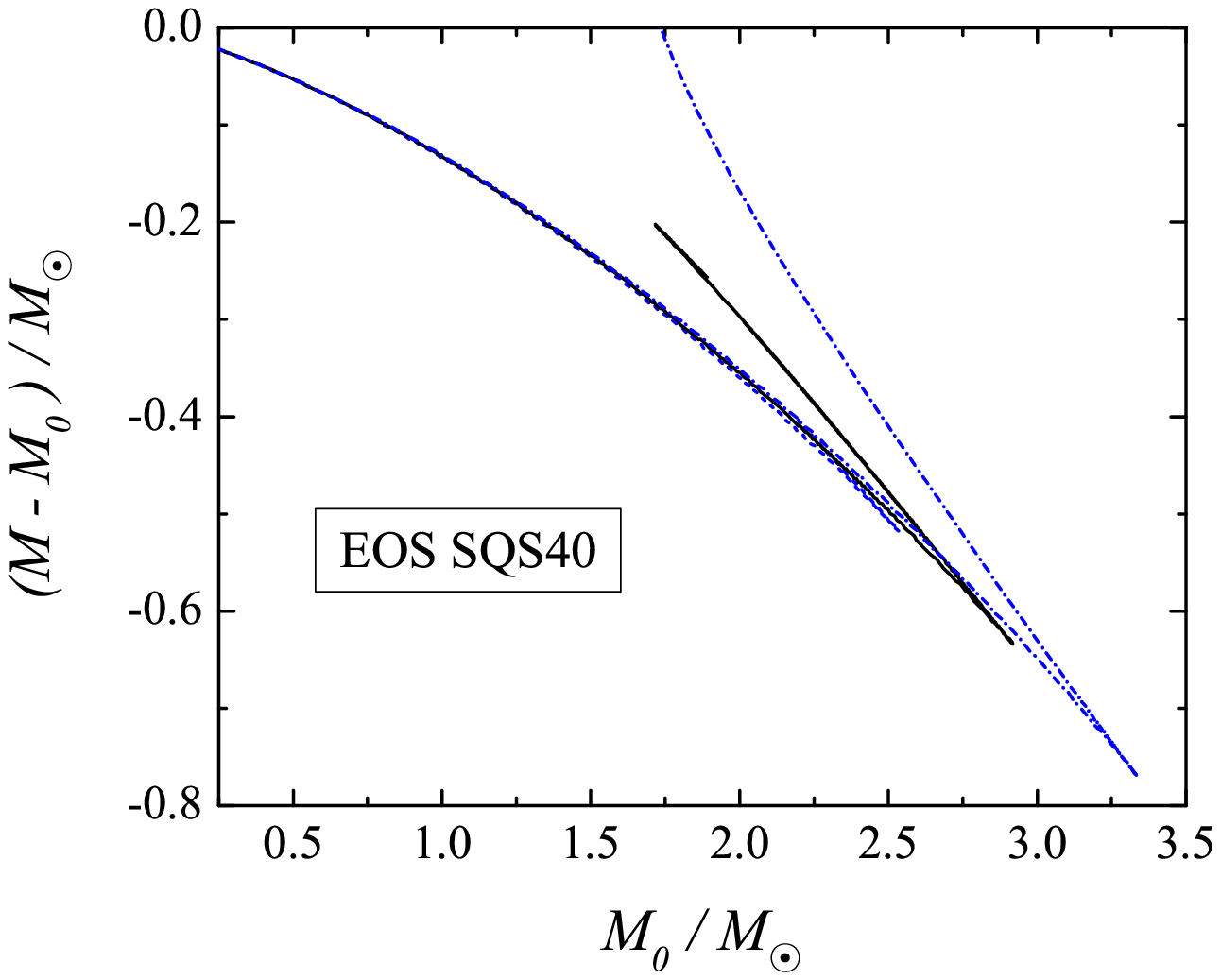}
	\caption{The binding energy $M-M_0$ as a function of the baryon mass for two equations of state -- the hadronic matter APR4 EOR (left panel), and the strange quark star SQS40 EOS (right panel). The results are plotted only for three values of $\alpha$ in order to have a better resolution.}
	\label{fig:BindingEn}
\end{figure}

The mass as a function of radius is presented in Fig. \ref{fig:MR} for  a representative hardonic matter  and strange star EOS and for several values of the parameter $\alpha$. Clearly, the $\alpha=0$ case is equivalent to pure general relativity. The maximum value of $\alpha$ we consider is $\alpha=3$ in agreement with the constraint discussed above.  The overall behavior of the $M(R)$ dependence is qualitative very similar to GR and differences exist only for models located far beyond the maximum of the mass that are considered unstable and we will not pay special attention to them. As one can see, negative (positive) $\alpha$ lead to a decrease (increase) of the neutron star radius and the maximum mass for the corresponding EOS. For the considered values of $\alpha$ and for the hadronic APR4 EOS, the decease (increase) of the maximum mass for a given equation of state is of the order of 20\% that is more or less within the uncertainty we have in the nuclear matter equation of state. The radius of a medium mass neutron star (e.g. $M=1.4M_\odot$) increases (decreases) by up to roughly 5\% for positive (negative) $\alpha$. The qualitative behavior of the mass of radius dependence is qualitatively similar for strange stars with the following main difference. As Fig. \ref{fig:BindingEn} (right panel) shows, the strange stars are almost indistinguishable from GR for small masses and we have larger deviations only close to the maximum mass. Similar behavior is observed for other alternative theories of gravity as well \cite{Staykov:2014mwa}. The changes in the maximum mass of EGB strange stars compared to GR models is more moderate that the hadronic matter EOS case and it is of the order of 10-15\%.

The stability of the obtained solutions can be roughly judged on the basis of the binding energy plotted in Fig. \ref{fig:BindingEn} as a function of the compact star baryon mass. Similarly to the pure general relativistic case, a cusp is observed at the maximum of the mass. This signals the appearance of instability for EGB compact stars beyond the maximum of the mass. Even though only a few representative cases are plotted in Fig. \ref{fig:BindingEn}, the same qualitative behavior is observed also for other values of $\alpha$ and other EOS.

Given the uncertainties in the determination of the neutron star radius \cite{Ozel:2016oaf}--\cite{Miller:2019cac} and the relatively small deviation caused by the EGB gravity, it is nearly impossible at present to constraint the theory on the basis of the neutron star radius observations. We can perform better, though, if we consider the mass of the compact star. As it is well established, the maximum neutron star mass $M_{\rm max}$ for a given equation of state should reach above the two solar mass threshold \cite{Demorest:2010bx,Antoniadis:2013pzd}. Since negative $\alpha$ lead to a decrease of $M_{\rm max}$ one can impose the requirement that $\alpha$ should be chosen in such a way that $M_{\rm max} > 2 M_\odot$. Clearly, such a requirement is EOS dependent. What one can do, though, is to assume that the true nuclear matter EOS is amongst the modern realistic EOS and from there see what is the minimum $\alpha$ that can still produce two solar mass neutron star. We will limit ourselves to the set of EOS considered in \cite{Read:2008iy} that is quite vast and covers a wide range of possibilities. Currently the observations favor EOS reaching slightly above two solar masses and having smaller radii for intermediate mass neutron stars \cite{Ozel:2016oaf}--\cite{Antoniadis:2013pzd}, but the interpretation of these observations is done assuming GR as the underlying theory of gravity and it is still unclear how the estimates of the neutron star mass and especially radius will change if me modify Einstein's theory. This is one of our main motivation behind considering a wider range of EOS some of which are outside the preferred radii range. Moreover, in this way a more robust lower limit on $\alpha$ can be imposed.

\begin{figure}
	\includegraphics[width=0.45\textwidth]{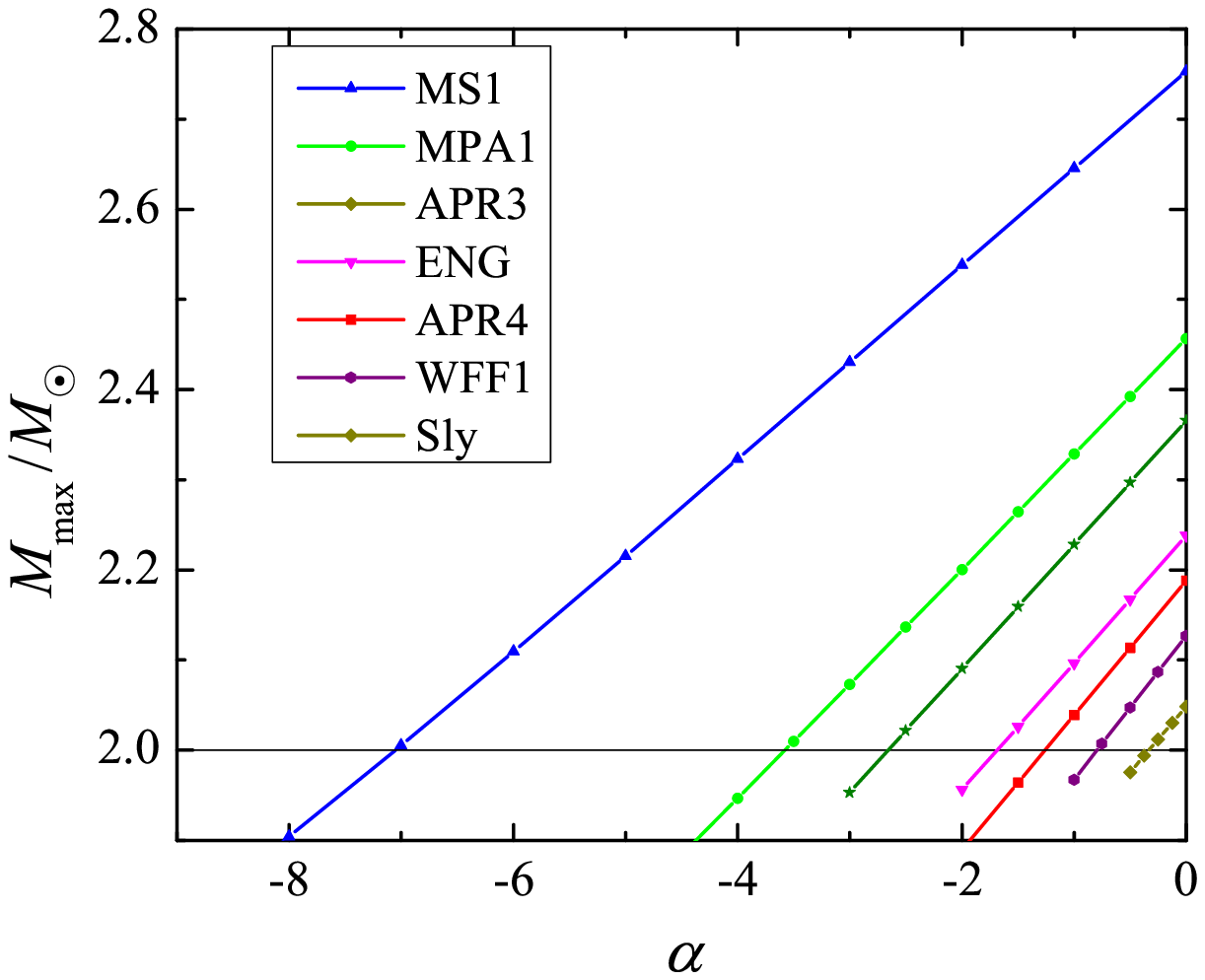}
	\includegraphics[width=0.45\textwidth]{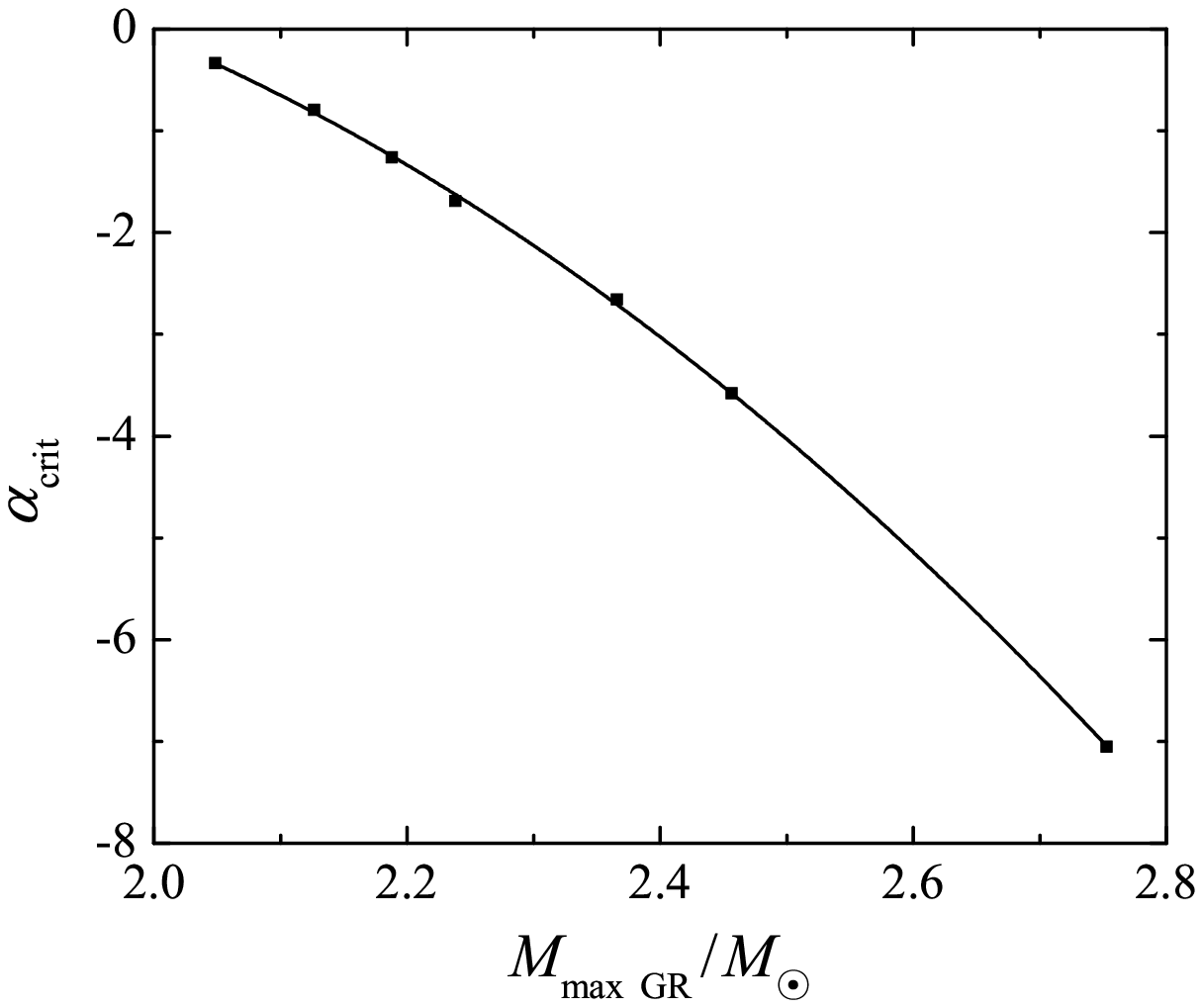}
	\caption{(left panel) The maximum mass $M_{\rm max}$ as a function of $\alpha$ for several EOS. (right panel) The critical $\alpha_{\rm crit}$ for which $M_{\rm max}$ is exactly equal to two solar masses, as a function of the maximum neutrons star mass in the GR  $\alpha=0$ limit (denoted by $M_{\rm max\; GR}$), for the corresponding EOS. Different points in the right figure correspond to the different EOS used in the left figure. The points are fitted with a second order polynomial. }
	\label{fig:Mmax_constraint}
\end{figure}

The maximum mass $M_{\rm max}$ as a function of $\alpha$ is plotted in Fig. \ref{fig:Mmax_constraint} for several EOS reaching above the two solar mass barrier. As one can notice, for the considered range of parameters, $M_{\rm max}$ is a monotonic function of $\alpha$ for a given EOS and  larger $M_{\rm max}$ in the GR $\alpha=0$ limit (denoted in the right panel of the figure by $M_{\rm max\; GR}$) lead to smaller value of the critical $\alpha_{\rm crit}$ for which $M_{\rm max}$ is equal to two solar masses. The EOS producing the most massive models is MS1\footnote{Up to our knowledge, as far as modern hadronic EOS are concerned, the maximum mass of the MS1 EOS is among the largest. Moreover this is the EOS with largest $M_{\rm max}$ in \cite{Read:2008iy}.} and therefore this EOS should be used in order to impose constraints on negative $\alpha$. The results show that $M_{\rm max}>2M_\odot$ for $\alpha\gtrsim -7$ in our dimensional units that corresponds to $\sqrt{|\alpha|}\lesssim 3.9 {\rm km}$ in dimensional units. 

Even though the MS1 EOS is quite extreme representative, our knowledge about the behavior of matter at very high densities is very limited and theoretically  EOS with larger $M_{\rm max}$ might be still allow. If one plots $\alpha_{\rm crit}$  as a function of the maximum mass in the GR limit $M_{\rm max\; GR}$ for a given EOS (right panel of Fig. \ref{fig:Mmax_constraint}), one can notice that the points for different EOS can be fitter very well with a second order polynomial and the exact form of this polynomial that we obtained from our results is
\begin{equation}
\alpha_{\rm crit} = -10.90 + 16.08 M_{\rm max} - 5.33 M_{\rm max}^2.
\end{equation}
Using this fit one can easily set limit on $\alpha_{\rm crit}$ for an arbitrary hadronic matter EOS. We should note, though, that the strange quark stars do not follow this nice polynomial behavior. It is difficult theoretically, though, to push the $M_{\rm max\; GR}$ for such EOS far beyond the two solar mass barrier and therefore, it is not expected that any realistic strange quark star EOS will give a stronger limit than the $\sqrt{|\alpha|}\lesssim 3.9 {\rm km}$ discussed above.

\section{Conclusions}
In the present paper we have studied compact stars in 4D Einstein-Gauss-Bonnet gravity which propagates only the massless graviton and bypasses the well-known Lovelock's theorem. The dimensionally reduced field equations governing the equilibrium compact star structure were derived and they were solved  analytically in the case of constant energy density, and numerically for realistic EOS.

The results show that the behavior of the equilibrium compact stars is very similar to the GR case and the mass and radius increases (decreases) for positive (negative) values of the Gauss-Bonnet coupling parameter $\alpha$. In our studies we have concentrated both on hadronic matter EOS and strange quark star EOS. The main difference between the two classes is that the deviation from GR is almost negligible for lower and intermediate mass strange stars while the differences between EGB and GR neutron stars described by a standard hadronic matter EOS can be significant in the whole mass range.  On the basis of the behavior of the binding energy it is expected that, similar to GR, the stability of the EGB solutions is lost at the maximum of the mass. 

We have discussed in detail the observational constraints on the parameter $\alpha$. While the existence of stellar mass black holes can constrain the  values of $\alpha$, namely we have that $\sqrt{\alpha}\lesssim 2.6 {\rm km}$  for $\alpha>0$, on can put better constraints on the negative $\alpha$ using the constructed neutron star models. Since negative $\alpha$ reduces the maximum mass of a compact star for a given equation of state, one can simply require that  $\alpha$ is chosen is such a way that this equation of state permits the existence of a two solar mass compact star. Even though this requirement depends heavily on the particular EOS one considers, if we limit ourselves to a set of moder realistic hadronic EOS, one can conclude that only when $\sqrt{|\alpha|}\lesssim 3.9 {\rm km}$  (for $\alpha<0$) one can still produce a two solar mass model for at least one of the considered EOS. We give as well a general formula for the threshold $\alpha$ which allows the existence of two solar mass neutron star for a given EOS as a function of the  corresponding GR maximum mass, which allows to modify the above mentioned limit for an arbitrary hadronic matter EOS.

\section*{Acknowledgements}
DD acknowledges financial support via an Emmy Noether Research Group funded by the German Research Foundation (DFG) under grant
no. DO 1771/1-1. DD is indebted to the Baden-Wurttemberg Stiftung for the financial support of this research project by the Eliteprogramme for Postdocs.  SY would like to thank the University of Tuebingen for the financial support.  
The partial support by the Bulgarian NSF Grant DCOST 01/6 and the  Networking support by the COST Actions  CA16104 and CA16214 are also gratefully acknowledged.


\end{document}